\documentstyle[12pt,epsf]{article}
\newcommand{\be}{\begin{equation}}
\newcommand{\ee}{\end{equation}}
\newcommand{\bea}{\begin{eqnarray}}
\newcommand{\eea}{\end{eqnarray}}
\newcommand{\bt}{\begin{tabular}}
\newcommand{\et}{\end{tabular}}
\newcommand{\ba}{\begin{array}}
\newcommand{\ea}{\end{array}}
\newcommand{\ov}{\overline}
\newcommand{\rrt}{\rightarrow}

\def\ane{\hbox{$\nu_e$ }}
\def\nm{\hbox{$\nu_\mu$ }}
\def\nt{\hbox{$\nu_\tau$ }}
\def\dm2{\hbox{$\Delta m^2$ }}
\def\th{\hbox{$\theta$ }}
\def\s2t{\hbox{$\sin (2 \theta$) }}
\def\ss2t{\hbox{$\sin^2 (2 \theta)$ }}
\def\c2t{\hbox{$\cos (2 \theta)$ }}
\def\cc2t{\hbox{$\cos^2 (2 \theta)$ }}

\def\emdm2{\hbox{$\Delta m^2_{e \mu}$ }}

\def\ems2t{\hbox{$\sin (2 \theta_{e \mu})$ }}
\def\emss2t{\hbox{$\sin^2 (2 \theta_{e \mu})$ }}
\def\emc2t{\hbox{$\cos (2 \theta_{e \mu})$ }}
\def\emcc2t{\hbox{$\cos^2(2 \theta_{e \mu})$ }}

\def\etdm2{\hbox{$\Delta m^2_{e \tau}$ }}

\def\ets2t{\hbox{$\sin(2 \theta_{e \tau})$ }}
\def\etss2t{\hbox{$\sin^2 (2 \theta_{e \tau})$ }}
\def\etc2t{\hbox{$\cos(2 \theta_{e \tau})$ }}
\def\etcc2t{\hbox{$\cos^2 (2 \theta_{e \tau})$ }}

\def\ene{\hbox{$\nu_e \rightarrow \nu_{e}$ }}
\def\emu{\hbox{$\nu_e \rightarrow \nu_{\mu}$ }}
\def\etau{\hbox{$\nu_e \rightarrow \nu_{\tau}$ }}
\def\noe{\hbox{$n^0_{\nu_e}$ }}
\def\nom{\hbox{$n^0_{\nu_{\mu}}$ }}
\def\anot{\hbox{$n^0_{\nu_{\tau}}$ }}
\def\nox{\hbox{$n^0_{\nu_x}$ }}
\def\nne{\hbox{$n_{\nu_e}$ }}
\def\nnm{\hbox{$n_{\nu_{\mu}}$ }}
\def\nnt{\hbox{$n_{\nu_{\tau}}$ }}

\begin{document}
\setcounter{page}{1}
\thispagestyle{empty}
\baselineskip=20pt
\hfill{
\begin{tabular}{l}
DSF$-$26/96 \\
hep-ph/9607226
\end{tabular}}

\bigskip

\begin{center}
{\Large \bf Supernova Neutrino Energy Spectra and the MSW Effect}
\end{center}

\vspace{1cm}

\begin{center}
{\large F. Buccella$^{1}$, S. Esposito$^{1}$, C. Gualdi$^{2}$ and
G. Miele$^{1}$} 
\end{center}

\vspace{1cm}

\normalsize
\noindent
{\it $^{1}$ Dipartimento di Scienze Fisiche, Universit\`a di Napoli 
''Federico II'', and INFN, Sezione di Napoli, Mostra D'Oltremare Pad. 20, 
I-80125 Napoli, Italy.}

\bigskip

\noindent
{\it $^{2}$ Dipartimento di Fisica, Universit\`a di Ferrara,
and INFN, Sezione di Ferrara, Via del Paradiso 12, I-44100 Ferrara,
Italy.}

\vspace{1cm}

\begin{abstract}
The distortions in the thermal energy spectra for neutrinos produced in
a supernova when a resonant oscillation, MSW effect, occurs are determined.
In order to show this effect for some relevant and representative examples
of unified gauge models, we have chosen $SO(10)$, and  $SU(5)_{SUSY}$, 
$SO(10)_{SUSY}$ with a particular scheme for fermion masses (DHR model).
The analysis has been performed for two choices of neutrinos parameters, 
predicted by the above models, and capable to explain 
the solar neutrino problem. In both cases one observes a strong 
distortion in the electron neutrino energy spectrum. This effect,
computed for a wide range of $SO(10)_{SUSY}$ models has produced the same 
results of the previous supersymmetric ones.
\end{abstract}

\vspace{1truecm}
\noindent
PACS: 13.15.+g; 14.60.Pq

\vspace{2truecm}

\noindent
e-mail: sesposito@na.infn.it; miele@na.infn.it

\normalsize
\newpage

\section{Introduction}
\indent
The long-base neutrinos physics certainly represents a suitable arena
to reveal the presence of fundamental effects, which due to their
weakness would be at the moment out of the range of accelerators
experiments. In particular, the explanation of the so-called solar
neutrino problem \cite{SNP} seems to suggest the presence of 
non-vanishing neutrino masses, and of a corresponding mixing \`a la
Cabibbo--Kobayashi--Maskawa also in the leptonic sector of the
standard model. Interestingly, since similar predictions are naturally
obtained in the framework of unified gauge models for fundamental
interactions, one can think to use the neutrino measurements to test
the validity of such models \cite{SO10,SO10bis}. 

In general, if one assumes massive neutrinos, the flavour eigenstates
do not coincide with the mass eigenstates, and thus the phenomenon of
neutrino oscillations can arise \cite{Pontecorvo}. In fact, neutrinos
are produced by weak interactions in flavour eigenstates, whereas
they propagates in space as energy eigenstates (mass eigenstates), 
and these two basis do
not coincide. Hence, there is a non-vanishing probability that
starting with neutrinos of given flavour, at a certain distance from
the source one can detect a neutrino of different flavour. Unfortunately, 
for solar neutrinos, this simple
oscillation mechanism (vacuum oscillations) is almost independent of
the neutrinos energy, and thus it hardly reproduces the experimental 
defect in the $\nu_{e}$ emission, which seems to show a quite strong 
energy dependence \cite{SNP}. A possible solution for the solar neutrino
problem is then provided by the so-called {\it resonant oscillation}  
mechanism, the Mikheyev--Smirnov--Wolfenstein effect (MSW) \cite{MSW}.

According to the MSW mechanism, 
when neutrinos propagate in a medium rich of electrons, neutrino 
oscillations can receive a resonant enhancement due to the different 
interactions experienced by $\nu_{e}$'s  (through both charged and neutral 
current) with respect to $\nu_{\mu}$'s and $\nu_{\tau}$'s (through only 
neutral current) with the medium itself. 

Let us consider for simplicity only two neutrino species, $\nu_{e}$ and
$\nu_{x}$ with $x=\mu$ or $\tau$. In this case if we denote with $E$ 
the neutrino energy, the enhanced oscillation occurs only 
if the following {\it resonance condition} is satisfied \cite{MSW}
\be
\frac{\dm2}{2E} \, \c2t \; = \; \sqrt{2} \, G_F \, N_e~~~,
\label{21}
\ee
where \dm2 is the squared masses difference between the two neutrinos, 
\th the vacuum mixing angle, $N_e$ the electron number density 
of the medium, and $G_F$ the Fermi coupling constant. 

As one can see from (\ref{21}), the resonance condition
if applied to astrophysical frameworks, would allow 
to scan very low values of \dm2 and $\theta$ by using quite 
low energy neutrinos, since very high densities can be achieved 
for large regions. One of
these situations is  certainly provided during a supernova explosion.

In this paper, we study how the MSW effect modifies the neutrinos energy 
spectra coming out from a supernova. The predictions are obtained 
for a wide range of unified gauge models. In particular, we have analyzed 
the $SO(10)$ predictions, as the most relevant example of non supersymmetric 
unified model still compatible with the experimental bounds on proton 
lifetime \cite{SO10,SO10bis}. Furthermore, for the sake of 
completeness, we have also considered 
$SU(5)$ and $SO(10)$ supersymmetric unified models 
\cite{DHR,SO10susy,SO10susybis}.

The paper is organized as follows. Section 2 is devoted to the study
of the MSW effect for medium with varying density. In particular, the
relevant quantities are defined and the expressions for neutrino
spectra are obtained. In Section 3 we review the main predictions for
the neutrino physics corresponding to the  still viable unified gauge
models, supersymmetric and not, and for these theories we compute the 
neutrinos spectra. Then in section 4 we give our conclusions and remarks. 

\section{MSW effect in supernov\ae} 

In media with varying density the probability of neutrino conversion 
depends on the rate of such a variation. Within the 
adiabatic approximation \cite{Adiabatic}, namely under the assumption
that the medium density crossed by neutrinos changes very 
slowly, one predicts an high rate of conversion. Quantitatively, this 
approximation is valid when the {\it adiabaticity parameter}
\be
\gamma \; \equiv \; \frac{\dm2}{2E} \, \frac{\ss2t}{\c2t} \, \left| 
\frac{1}{N_e} \, \frac{d N_e}{d r} \right| ^{-1}_{res}~~~,
\label{22}
\ee
computed at the resonance point, is greater than 1. In the opposite 
case, the survival probability would be of the order of unity, so that no 
appreciable conversion can be obtained. For neutrinos propagating in a 
given medium it is useful to introduce the adiabatic and non-adiabatic 
threshold energies
\begin{eqnarray}
E_A \; &=& \; \frac{\dm2 \, \c2t}{2 \sqrt{2} \, G_F \, N_e^{prod}}~~~,
\label{23}\\
E_{NA} \; &=& \; \frac{\pi \, \dm2 \, \ss2t}{4 \, \c2t} \, \left| 
\frac{1}{N_e} \, \frac{d N_e}{d r} \right| ^{-1}_{res}~~~,
\label{24}
\end{eqnarray}
where $N_e^{prod}$ is the electron density at the neutrinos production
point\footnote{For simplicity, one can assume that this point essentially 
coincides with the neutrinosphere.}.
The former indicates the transition between no matter effects and 
resonant conversion, described by the adiabatic approximation, while 
the latter represents the energy scale for the transition between the 
adiabatic and non-adiabatic regions.

The density range in a supernova is very large; it goes from 
$\sim 10^{-5} g/cm^3$ in the external envelope up to 
$\sim 10^{15} g/cm^3$ in the dense core. For our purposes, as a reasonable 
radial density profile one can use the following scaling law 
\cite{Adiabatic,densita}
\be
\rho \; \simeq \; \rho_0 \, \left( \frac{R_0}{r} \right) ^3~~~,
\label{213}
\ee
with $\rho_0 \simeq 3.5 \times 10^{10} g/cm^3$, and $R_0 \simeq 1.02 
\times 10^7 cm$. The quantity $R_{0}$ defines the so-called 
{\it neutrinosphere}, which represents the bounding surface 
of the region in which neutrinos of a given flavour are in
thermal equilibrium (eq. (\ref{213}) is approximatively valid for
$r \geq R_0$).\\
Outside the neutrinosphere, the electron fraction number $Y_e$ 
can be assumed almost constant with
respect to the radius, and thus the value $Y_e = 0.42$
can be used. 
Since inside the neutrinosphere the resonance conditions
are not satisfied, for a wide range of unified models, we can 
assume in first approximation that on this surface the neutrinos 
are emitted with an almost Fermi-Dirac distribution, and with the
different flavors equally populated. However, flavour changing 
effects occurring inside the neutrinosphere have been proposed 
in literature \cite{insideneutrinosphere}, but in any case they 
would represent further corrections with respect to the bulk
phenomenon we want to describe.

Consequently,  
from (\ref{21}) it follows that neutrinos diffusing out the neutrinosphere
experience two MSW resonances: the first resonance, between \ane 
and \nt, at a density of about $10^8~ g/cm^3$ and the second one, 
between \ane and \nm, at about $10^2~ g/cm^3$. These values refer to  
average neutrino energies $\overline{E}_{\nu_e} \simeq 10 \, MeV$, 
$\overline{E}_{\nu_{\mu}, \nu_{\tau}} \simeq 20 \, MeV$, 
and to the indicative values $\Delta m^2_{e \mu} \, \simeq 10^{-6} \, eV^2$,
$\Delta m^2_{e \tau} \, \simeq 25 \, eV^2$ for the squared masses
differences.
The evolution of the different neutrino flavors can be simply
summarized by the following diagrams
\be
\ane \, \left\{
\ba{l}
-----> \; \nt \; ----------> \; \nt \\
----------> \; \nm \; ~ -----> \; \nm \\
------------------~>  \; \ane
\ea \right. 
\nonumber
\label{d1}
\ee
\be
\nm \, \left\{
\ba{l}
----------> \; \ane \; ~~ -----> \; \ane \\
------------------~> \; \nm
\ea \right.
\nonumber
\label{d2}
\ee
\be
\nt \, \left\{
\ba{l}
-----> \; \ane \; ----------> \; \ane \\
~~~~~~~~~~~~~~~~~~~ \, ~~~~~   \; --> \; \nm \; -----> \; \nm \\
------------------~> \; \nt
\ea \right.
\nonumber
\label{d3}
\ee
Electron-neutrinos may oscillate into \nt at the first resonance or 
into \nm at the second resonance, whereas muon-neutrinos can oscillate only 
into \ane at the second resonance. Finally, tau-neutrinos may oscillate 
into \ane at the first resonance and the latter can further oscillate into 
\nm at the second resonance. 

To study the evolution of neutrino flavors 
it is crucial to know how these resonance regions are 
crossed, namely if adiabatically or non-adiabatically. 

\subsection{The \ane survival probability and spectra distortion}

In order to obtain the energy spectra of the emitted neutrinos 
from a supernova, let us focus our attention on neutrinos produced in 
the {\it thermal phase}, which diffuse out the neutrinosphere with a spectrum 
approximatively given by \cite{densita,spettro}\footnote{
Note that we assume neutrino masses negligible with respect to $T_i$.}
\be
n_{\nu_i}(E) \; \simeq \; \frac{0.5546}{T_i^3} \, E^2 \left[
1+\exp\left(\frac{E}{T_i}\right) \right]^{-1}~~~,
\label{3.1}
\ee
where the index $i$ denotes the particular neutrino specie.
Since the production and scattering cross-sections for electron-neutrinos
are larger than for the other flavors, \ane are produced a bit 
more copiously with respect to the other ones. Thus, their neutrinosphere 
is larger than that for \nm, \nt. This implies that the 
temperature $T_i$ in (\ref{3.1}) for the \ane-sphere is lower 
than the one of \nm, \nt. 
Furthermore, since \nm and \nt are produced and scatter on the 
surrounding matter only through neutral currents, they have identical 
spectra. Obviously, since \ane, \nm, \nt and $\ov{\nu}_e$, 
$\ov{\nu}_{\mu}$, $\ov{\nu}_{\tau}$ are produced in pairs, 
the magnitude of neutrino and 
antineutrino fluxes are equal for each flavour. For the temperature of 
\ane and \nm, \nt neutrinosphere we adopt the typical values 
\cite{densita}, \cite{Adiabatic}
\begin{eqnarray}
T_{e} \; &\simeq& \; 3 \, MeV~~~,\label{32}\\
T_{\mu}&=&T_{\tau} \; \simeq \; 6 \, MeV~~~.\label{33}
\end{eqnarray}
By indicating with \noe and \nox $\equiv$ \nom = \anot the spectra of \ane 
and \nm , \nt at the relative neutrinospheres, according to the 
oscillation scheme previously outlined, we have that the spectra of 
the different neutrino flavors emerging from a supernova are
\begin{eqnarray}
\nne \; &=& \; P( \ene ) \, \noe \; + \; \left[ P( \nm 
\rightarrow \ane ) \, + \; P( \nt \rightarrow \ane ) \right] \, \nox~~~, 
\label{34}\\
\nnm \; &=& \; P( \emu ) \, \noe \; + \; \left[ P( \nm \rrt \nm ) \, + 
\, P( \nt \rrt \ane \rrt \nm ) \right] \, \nox~~~,\label{35}\\
\nnt \; &=& \; P( \etau ) \, \noe \; + \; P( \nt \rrt \nt ) \, \nox~~~,
\label{36}
\end{eqnarray}
where $P(\nu_{\alpha} \rightarrow \nu_{\beta})$ is the probability for 
the transition $\nu_{\alpha} \rightarrow \nu_{\beta}$. Note 
that, from the experimental point of view, due to the relative low energy 
of the neutrinos considered ($0 \div 50~ MeV$), one cannot distinguish 
between \nm and \nt, thus the only relevant quantities are \nne and 
\nnm + \nnt. Consequently, by using the unitarity conditions on the 
transition probabilities (see diagrams (\ref{d1})-(\ref{d3})) we obtain
\begin{eqnarray}
\nne \; &=& \; P( \ene ) \, \noe \; + \; \left[ 1 \, - \, P(\ene ) 
\right] \, \nox~~~,\label{37}\\
\nnm \, + \, \nnt \; &=& \; \left[ 1 \, - \, P( \ene ) \right] \, \noe 
\; + \; \left[ 1 \, + \, P( \ene ) \right] \, \nox~~~.
\label{38}
\end{eqnarray}
From the above equations it is evident that the only relevant quantity 
to be computed from the MSW theory is the probability that \ane survives to 
both \etau and \emu resonances. Since for a wide range of unified models
the two resonances are well separated we can write
\be
P( \ene ) \; \simeq \; P_{e \tau} ( \ene ) \, P_{e \mu} ( \ene )~~~,
\label{39}
\ee
where $P_{e \tau} ( \ene )$ and $P_{e \mu} ( \ene )$ are the survival
probabilities for \ane's to the first and second resonance. These 
quantities can be calculated by
using a simple two flavour approach. According to this we have 
\cite{Adiabatic}
\be
P_{e \mu} ( \ene ) \; = \; \frac{1}{2} \; + \; \left( \frac{1}{2} \, - 
\, P^c_{e \mu} \right) \cos (2 \theta_{e \mu}) \cos (2 \theta_m^{e \mu})~~~,
\label{310}
\ee
and an analogous form for $P_{e \tau} ( \ene )$. Here $\theta_m$ denotes the 
effective mixing angle in matter calculated at the neutrinosphere. It is 
given by
\be
\cos(2\theta_m) = \left\{
\ba{lcl}
\mbox{sign}(E_A  -  E) 
\left[1 + \tan^2(2 \theta)  \left( { E_A  \over E_A - E }\right)^2 
\right]^{-1/2} & & E {\neq} E_A \\
& & \\
0 & & E = E_A
\ea \right.
\label{311}
\ee
where $P^c$ is the probability of level crossing at the resonance. 
For the density scaling law (\ref{213}), it has the expression
\be
P^c  =  \exp\left[- \left( {E^{\ast}_{NA} \over E} \right) 
^{\frac{2}{3}} F(\theta) \right]~~~,
\label{312}
\ee
where the function $F(\theta)$ is approximatively given by 
\be
F(\theta) \simeq \left(1 \, - \tan^2(\theta)\right) \left\{ 1 + \frac{1}{3} 
\left[ \log\left(1 - \tan^2(\theta)\right) + 1 - \frac{1 + \tan^2(\theta)}
{\tan^2(\theta)} \log\left(1+\tan^2(\theta)\right) \right] \right\},
\ee
and $E^{\ast}_{NA}$ is the effective non-adiabatic threshold energy 
for density scaling as $1/r^3$
\be
\left( \frac{E^{\ast}_{NA}}{1 \, MeV} \right) ^{\frac{2}{3}}  \simeq 
 2.4 \times  10^6 ~\left( \frac{\dm2}{1 \, eV^2} \right) 
^{\frac{2}{3}} \, \frac{\ss2t}{[ \c2t ]^{\frac{4}{3}}} \, 
Y_e^{\frac{1}{3}}~~~.
\label{313}
\ee

\section{Supernova neutrino spectra for unified gauge models}

Let us apply the previous results to some relevant example of 
unified gauge models. In particular, we will consider the unified 
gauge model based on $SO(10)$ \cite{SO10,SO10bis} which among the 
non supersymmetric GUT theories, represents the most appealing
unification scheme. Furthermore, we will extend our analysis
to the SUSY version of $SU(5)$ and $SO(10)$ recently extensively
discussed in literature \cite{DHR,SO10susy,SO10susybis}.

\subsection{$SO(10)$ GUT models and neutrino predictions}

As already noted in the previous section, the MSW effect strongly depends 
on the neutrino mixing angle \th and on \dm2. For \emu transitions, however,
one can restrict the values of these quantities
by requiring that MSW explains the solar neutrino problem 
\cite{SNP}. This analysis, performed by one of the present authors 
\cite{Espo}
has given the following results
\begin{eqnarray}
\emdm2 & \simeq &~ 6  \times  10^{-6} ~ eV^2~~~,\label{25}\\
\sin^2(2 \theta_{e \mu}) & \simeq & 7  \times  10^{-3}~~~.
\label{26}
\end{eqnarray}
A value for the mass of \nm of the order of the square root
of the r.h.s. of Eq. (\ref{25}) (and a value for the mass of
\ane several order of magnitude smaller) is found in $SO(10)$
models with intermediate symmetry $SU(4)_{PS} (\mbox{or}~
SU(3)_C \otimes U(1)_{B-L}) \otimes SU(2)_{L} \otimes SU(2)_{R}$.
In general, a mass for \nt about $m_t^2/m_c^2$ larger than 
the one of \nm is expected \cite{SeeSaw}. 
This implies, provided $m_{\ane} << m_{\nm}$, 
\be
m_{\nt} \approx 35~eV~~~,
\label{27}
\ee
larger than $\sim 5~eV$ which is the value found if one identifies \nt 
as the hot component of dark matter \cite{Gelmini}. However,
the value given by Eq. (\ref{27}) is obtained by assuming equal 
Majorana masses for $\nu_{\mu R}$ and $\nu_{\tau R}$. Thus, 
the factor $7$ of discrepancy may be just ascribed to this ansatz.
For this reason one can safely take for $m_{\nt}$ the value $5~eV$.

By applying Eq. (\ref{23}) this value corresponds to an adiabatic 
threshold  for the $\nu_{e} \Leftrightarrow \nu_{\tau}$ transition 
in the matter of about $10^{-2}~MeV$ (since we expect small mixing angle 
we take 
$\cos(2\theta_{e\tau}) \simeq 1$). The efficiency in the conversion  
$\nu_{e} \Leftrightarrow \nu_{\tau}$ depends on the value 
of the mixing angle between \ane and \nt. To be more precise, it can be
large if the threshold of the non adiabatic region is larger than 
the adiabatic threshold , or small in the opposite case, if the adiabatic
region vanishes.\\
Let us denote with $\tilde{\theta}_{e \tau}$ the value of the mixing angle for 
which the two thresholds (\ref{23}) and (\ref{313}) coincide.
With the density scaling given by (\ref{213}) we get
\be
\sin^2(2\tilde{\theta}_{e \tau}) = 3.25 \times 10^{-9}~~~.
\label{28}
\ee
Therefore, for $\tilde{\theta}_{e \tau} > \theta_{e \tau}$, the MSW
conversion $\nu_{e} \Leftrightarrow \nu_{\tau}$ will be important,
otherwise it will be small. Then in the following analysis we will consider
the two values $\sin^2(2 \theta_{e \tau})=3.25~10^{-8}$ and
$3.25 \times 10^{-10}$, representatives of the two widely different 
regimes discussed above.

By using the above values for neutrino parameters we can apply the MSW theory 
to obtain the spectra of emitted neutrinos from a supernova. 
In particular, Figure 1 shows the adiabaticity parameter $\gamma$ 
for \emu (solid line) and \etau transitions
(dashed line), respectively, versus the neutrino energy. 
Note that, the dashed line corresponds to the larger value for
$\sin^2(2 \theta_{e \tau})$, since to the other choice just would
correspond a line two order of magnitude lower.

The plots show clearly that the \etau resonance is crossed much less 
adiabatically than \emu if values for $\sin^2(2 \theta_{e \tau})<<
3.25 \times 10^{-9}$ are considered. Hence in these cases, the transition \emu 
is expected to be more effective than \etau.

In Figures 2 and 3, 4 the $\nu_{e}$ survival probabilities, corresponding to
the resonant transitions $\nu_{e} \rightarrow \nu_{\mu}$ and
$\nu_{e} \rightarrow \nu_{\tau}$ respectively, are plotted versus the neutrino 
energy. From Figure 3 the extreme non-adiabatic nature of 
the \etau transition for $\sin^2(2 \theta_{e \tau}) \simeq 
3.25 \times 10^{-10}$ is clear. 

In terms of the survival probability one can calculate the 
modifications of neutrino energy spectra as induced by MSW 
oscillations. In Figures 5 and 6 are plotted \noe, \nne and \nom + \anot, 
\nnm + \nnt respectively as given by equations (\ref{37}), (\ref{38}).
In particular, the dashed and dotted lines represents the $SO(10)$ predictions
for the values $\sin^2(2 \theta_{e \tau})=3.25 \times 10^{-10}$ and
$3.25\times 10^{-8}$, respectively.

As one can see from these Figures, the \ane 
spectrum becomes harder toward higher energies (more easily detectable) 
while \nm, \nt emerge from a supernova with lower energies. The 
integral flux of \ane is also partially reduced of about $16\%$.

However, 
the possible values of the neutrino mixing angles in $SO(10)$ 
models with fermion couplings to a $\underline{126}$  and a complex 
$\underline{10}$ , as it is the case of the models discussed in
Ref. \cite{SO10}, has been widely discussed by Babu and Mohapatra
\cite{SO10bis}. They conclude that it is 
possible to get the $\nu_{e} \Leftrightarrow \nu_{\mu}$ MSW
mechanism advocated to explain the solar neutrino
experiments with $\sin^2(2\theta_{e \tau}) \approx 10^{-2}$, much
larger than the r.h.s. of Eq. (\ref{28}). In this case the 
$\nu_{e} \Leftrightarrow \nu_{\tau}$ transition in supernovae
is extremely efficient and thus the corresponding
$P_{e \tau} ( \ene )$ in Eq. (\ref{39}) is almost vanishing.
Thus, the expressions of the neutrino spectra, as can be seen from 
(\ref{34})-(\ref{36}), take the simple form
\begin{eqnarray}
\nne \; &=& \; \nox~~~,\label{28a}\\
\nnm \, &=& \; \nox~~~,\label{28b}\\
\nnt \; &=& \; \noe~~~.\label{28c}
\end{eqnarray}
It is interesting to observe that in this case the \nt and \ane 
distributions just interchange, whereas the \nm one remains 
unmodified. This result can be easily understood by observing that 
for such an efficient $\nu_{e} \Leftrightarrow \nu_{\tau}$ 
transition, all the final \ane come either from the former \nt,
not subsequently transformed in \nm, or from the initial
\nm just transformed in \ane. Thus, since the initial distributions 
of \nm and \nt are assumed equal, this yields to the result
(\ref{28a})-(\ref{28c}).

\subsection{$SU(5)_{SUSY}$ and $SO(10)_{SUSY}$ unified models and 
neutrino parameters}
As an example of a supersymmetric grand unified model, we consider 
the scheme discussed by Dimopoulos, Hall and Raby (DHR) \cite{DHR} which 
is based on 
the Georgi-Jarlskog texture \cite{GJ} for fermion mass matrices at the 
GUT scale. A peculiarity of the DHR model is that it can be realized 
both in $SU(5)_{SUSY}$ and in $SO(10)_{SUSY}$ GUT models (obviously, the 
Higgs multiplets in the two cases are different). The predictions for 
the neutrino mass ratios of this model are
\begin{eqnarray}
\frac{m_{\nm}}{m_{\ane}} & = & 9 ~ k^4 ~ \frac{m_{c}}{m_{t}}~~~,\\
\frac{m_{\nt}}{m_{\nm}} & = & \frac{1}{3 k^2} \left( \eta_3 V_{cb}^q 
\right)^{-2}~~~,
\end{eqnarray}
where $k = 1$ or $- 1/3$ (depending on particular 
choices in the DHR framework, see \cite{DHR}), and $\eta_3=
\eta_c~(V_{cb}^q)^2~m_t/m_c$, $\eta_c$ being the 
QCD renormalization enhancement factor of the charm mass between 
$m_{t}$ and $m_{c}$. The three leptonic mixing angles 
($\theta^{\prime}_1$, $\theta^{\prime}_2$, $\theta^{\prime}_3$) are 
expressed in terms of the three quark 
mixing angles ($\theta_1$, $\theta_2$, $\theta_3$) by
\begin{eqnarray}
\sin(\theta^{\prime}_1)&=& - \frac{1}{3} ~\sin(\theta_1)~~~,\\
\sin(\theta^{\prime}_2)&=&  \frac{1}{3 k^2}~\sin(\theta_2)~~~,\\
\sin(\theta^{\prime}_3)&=& 2~ k~ \eta_3~\sin(\theta_3)~~~.
\end{eqnarray}
In particular,
\begin{eqnarray}
\theta_{e \mu}  &\simeq&  \sqrt{ \frac{m_{e}}{m_{\mu}} + 
\frac{m_{\ane}}{m_{\nm}} - 2~\sqrt{\frac{m_{e}}{m_{\mu}}
\frac{m_{\ane}}{m_{\nm}}}~\cos(\phi)}~~~,\\
\theta_{e \tau} &\simeq& \sin(\theta^{\prime}_1)~ \sin(\theta^{\prime}_3)
\end{eqnarray}
where $\phi$ is the phase of the leptonic CKM matrix.\\
Inserting in the above formulas the known experimental parameters, one 
obtains \cite{DHR}
\begin{eqnarray}
\frac{m_{\nm}}{m_{\ane}} &= & \left\{ 
\ba{lcl}
3100 \pm 1000 & & \mbox{I}\\
38 \pm 12 & & \mbox{II}
\ea \right.~~~,\\
\frac{m_{\nt}}{m_{\nm}} &= & \left\{ 
\ba{lcl}
208 \pm 42 & & \mbox{I}\\
1870 \pm 370 & & \mbox{II}
\ea \right.~~~,\\
\sin^2(2 \theta_{e \mu})& = & \left\{ 
\ba{lcl}
(1.7 \pm 0.2) 10^{-2} & & \mbox{I}\\
(9.0 \pm 4.3) 10^{-2} & & \mbox{II}
\ea \right.~~~,\\
\sin^2(2 \theta_{e \tau})& = & \left\{ 
\ba{lcl}
(1.3 \pm 0.3) 10^{-6} & & \mbox{I}\\
(1.4 \pm 0.3) 10^{-7} & & \mbox{II}
\ea \right.~~~,
\end{eqnarray}
where I and II refer to the different values of $k$.\\
In this framework, as done in the previous section, one can fix the \emu mixing 
angle and squared masses difference to those values capable to
explain the solar neutrino problem via the MSW mechanism ($\emss2t \, = 
\, 6 \times 10^{-4} \div 2 \times 10^{-2}$, $\emdm2 \, = \, (4 \div 9) 
10^{-6} \, eV^2$ , see \cite{SNP}). This requirement selects the model 
I with the following squared masses differences and mixing angles
\begin{eqnarray}
\emdm2 &\simeq & 6 \times 10^{-6}~eV^2~~~,\\
\etdm2 &\simeq & 0.27 ~ eV^2~~~,\\
\emss2t &\simeq &1.7 \times 10^{-2}~~~,\\
\etss2t &\simeq & 1.3 \times 10^{-6}~~~.
\end{eqnarray}
In Figures 5 and 6 the predictions for the neutrino spectra
for the DHR model, which is compatible with both $SU(5)_{SUSY}$ 
and $SO(10)_{SUSY}$, are shown (dashed-dotted lines).

As a possible alternative scenario, Lee and Mohapatra \cite{SO10susybis}
have also considered the possibility of SUSY $SO(10)$, directly
broken at a scale $\sim 10^{16}~GeV$ to $SU(3) \otimes 
SU(2) \otimes U(1)$. In this scheme one expects Majorana masses 
for the $\nu_{R}$'s at the highest scale and it is the \nt 
the neutrino expected to have a mass around $10^{-3}~eV$. In
that case the mass of \nm is too small to cause a $\nu_{e}
\Leftrightarrow \nu_{\mu}$ MSW conversion in an experimentally
interesting range of neutrino energies. In this case, in supernova
one should have only the conversion $\nu_{e} \Leftrightarrow \nu_{\tau}$,
responsible for the solution of the solar neutrino problem.
The situation would be very similar, as long as the distortions of neutrino
spectrum are concerned, to a situation, where the MSW $\nu_{e} 
\Leftrightarrow \nu_{\mu}$ conversion is responsible for the solution 
of the solar neutrino problem and the mixing angle $\theta_{e \tau}$
is smaller than $\tilde{\theta}_{e \tau}$.

\section{Conclusions}
In this paper we have analyzed the distortions induced on the
thermal energy spectra of $\nu_{e}$ and 
$\nu_{\mu}+\nu_{\tau}$ emitted from a supernova due to
MSW effect. In order to study this phenomenon for some relevant
examples of unification scenarios, we consider as
possible choices of the neutrinos parameters, namely
masses and mixings, the ones corresponding to the $SO(10)$ gauge group
\cite{SO10,SO10bis}, and the $SU(5)_{SUSY}$, $SO(10)_{SUSY}$ unified models 
in the scheme DHR \cite{DHR}, respectively. Other different $SO(10)_{SUSY}$ 
models have been also considered \cite{SO10susy}, 
and interestingly, as far as the neutrino spectra predictions are concerned,
it yields to results almost equal to the previous supersymmetric ones.
Note that, in all cases \emdm2 and \emss2t have been chosen to be compatible 
with the explanation of the solar neutrino problem \cite{SNP}.

According to the typical values for the density distribution in a supernova, 
out of the neutrinosphere, the condition of resonance is satisfied for
the two flavour transitions $\nu_{e} \rightarrow \nu_{\mu}$ and 
$\nu_{e} \rightarrow \nu_{\tau}$, respectively. By computing the 
corresponding adiabaticity parameter $\gamma$, one observes that, at least for
$SO(10)$ GUT theories, the resonance $\nu_{e} \rightarrow \nu_{\tau}$
is crossed less adiabatically than $\nu_{e} \rightarrow \nu_{\mu}$
if one takes values $\sin^2(2 \theta_{e \tau})<<3.25 \times 10^{-9}$,
and thus the latter results the most efficient. The survival probabilities
for the electron-neutrinos corresponding to the above resonances are then
obtained. 

Finally, the distortion of the energy spectra are obtained for the $\nu_e$ and 
$\nu_{\mu}+\nu_{\tau}$, emitted from the supernova neutrinospheres in pure
thermal equilibrium distributions. Remarkably, for both models
$SO(10)$ and $SU(5)_{SUSY}$, $SO(10)_{SUSY}$ in DHR scheme,
the distortion in the $\nu_{e}$ 
distribution is quite relevant, and moves neutrinos from
the low energy region of the spectrum to the high energy sector.
As a result of this analysis, one observes
that the reduction of $\ane$ flux for $SO(10)$ and
for the DHR model is of the order $16\%$. Thus it is essentially model
independent. This feature can be easily understood by observing that
the relevant (since almost adiabatic) neutrino transition is
$\emu$, whose parameters are fixed from the MSW explanation for 
the solar neutrino problem.\\
The distortion in the \ane distribution becomes extreme when the
choice of neutrino parameteres discussed in the papers of Babu and 
Mohapatra \cite{SO10bis} is taken. In this case, in fact, since
the survival probability $P_{e \tau}(\nu_{e} \rightarrow \nu_{e})$
is almost vanishing, we have the exact interchange of \ane and \nt
distributions.

\newpage

\begin{figure}[]
\epsfysize=18cm
\epsfxsize=16cm
\epsffile{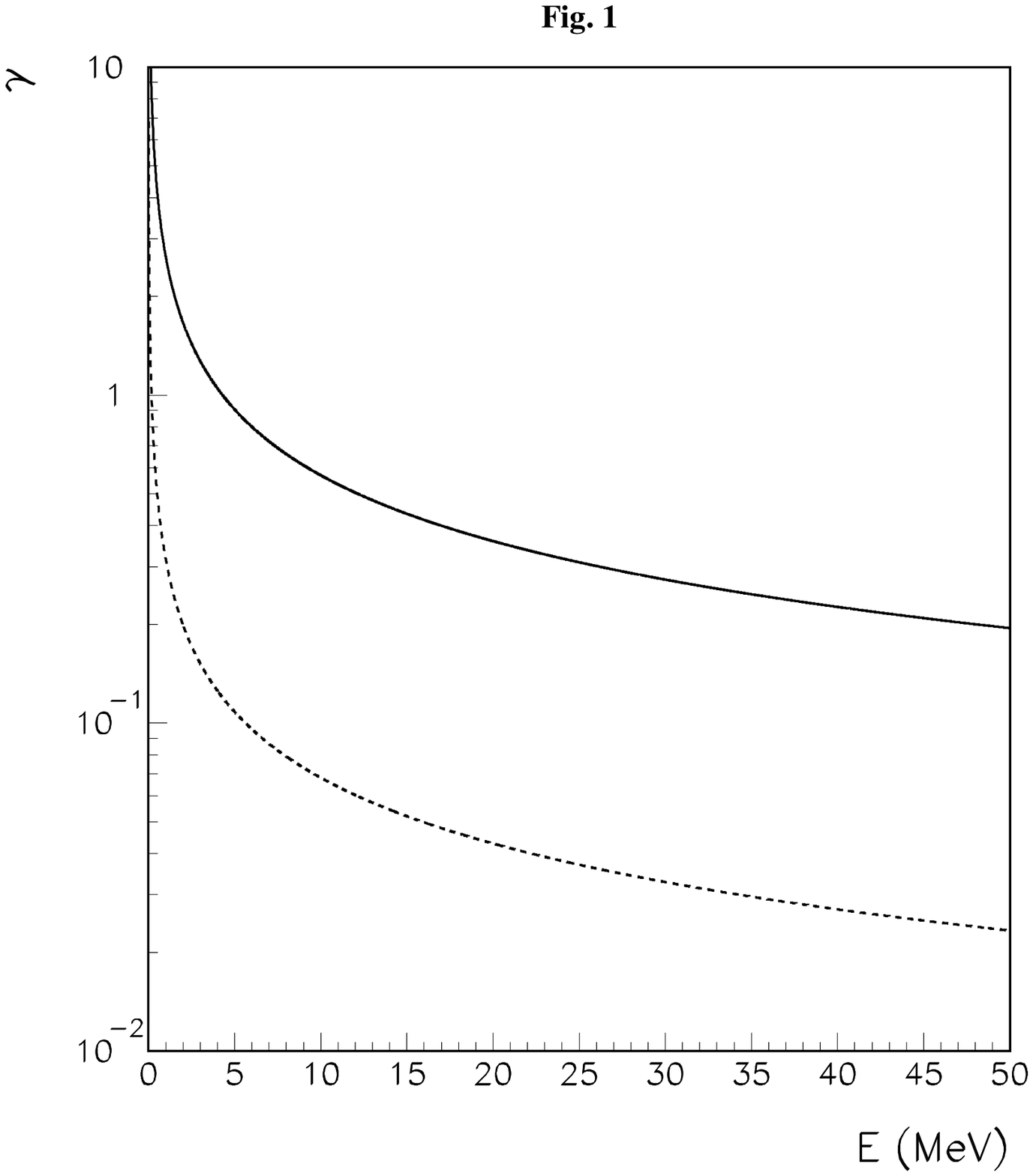}
\caption[]{The adiabaticity parameter $\gamma$ for \emu (solid line)
and for \etau transitions (dashed line) versus the neutrino energy
is reported, for the $SO(10)$ unified model \cite{SO10bis}. Note that
for the $SO(10)$ case the line corresponds to $\sin^2(2 \theta_{e \tau})
=3.25~10^{-8}$.}
\end{figure} 

\newpage

\begin{figure}[]
\epsfysize=18cm
\epsfxsize=16cm
\epsffile{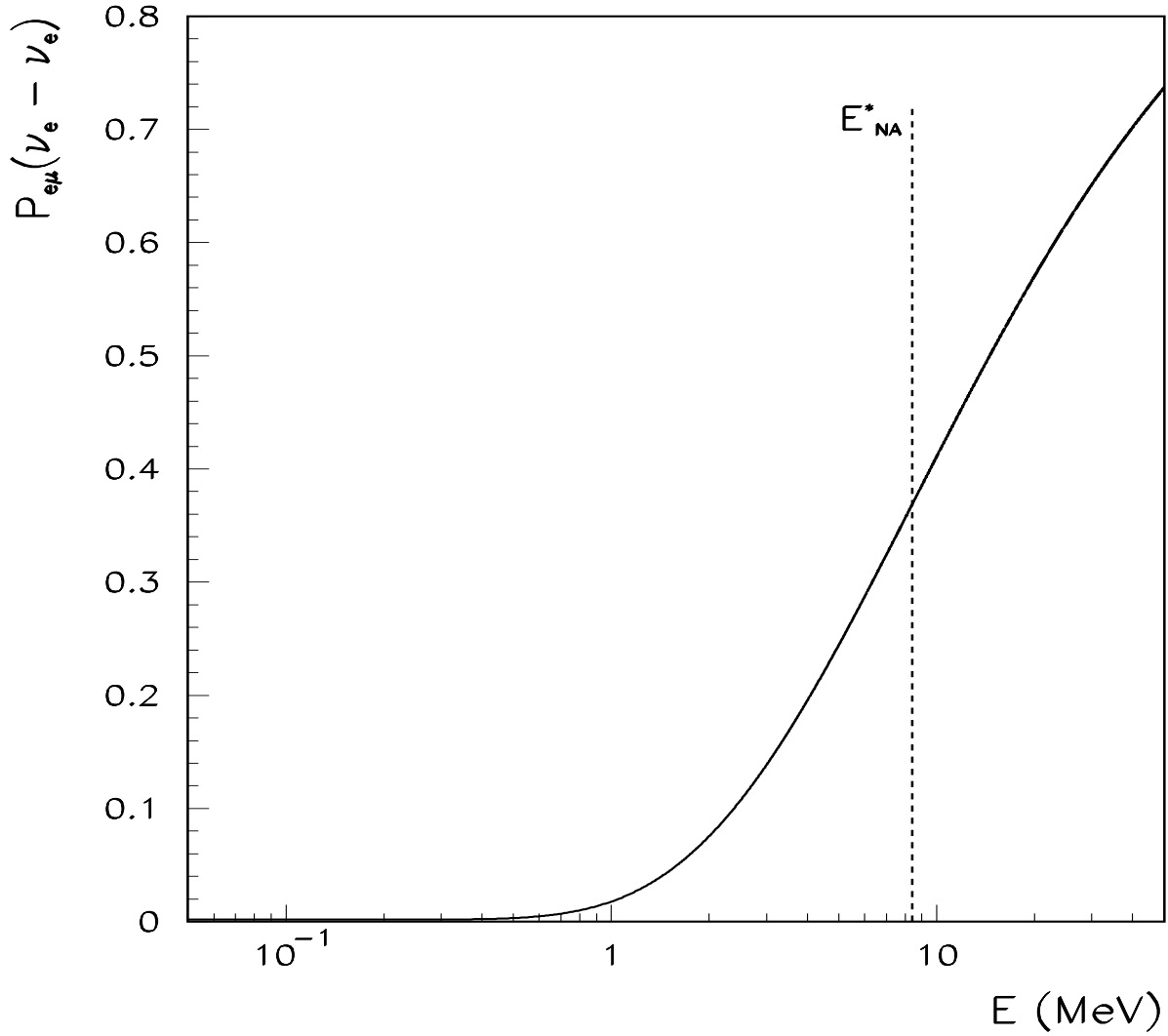}
\caption[]{The survival probability $P_{e \mu}(\nu_{e} \rightarrow \nu_{e})$
as function of the neutrino energy is shown for the $SO(10)$ \cite{SO10}.}
\end{figure} 

\newpage

\begin{figure}[]
\epsfysize=18cm
\epsfxsize=16cm
\epsffile{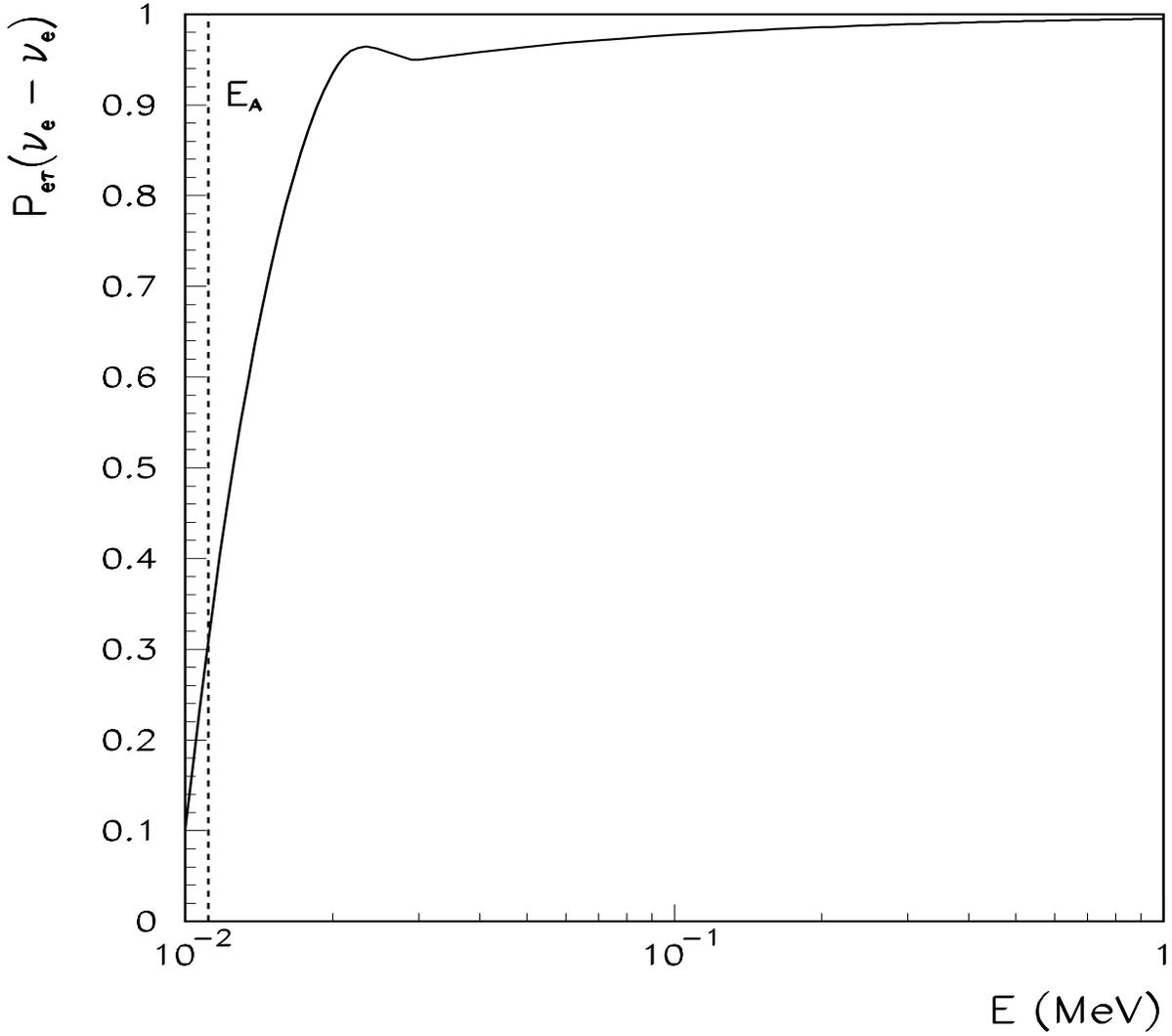}
\caption[]{The survival probability $P_{e \tau}(\nu_{e} \rightarrow \nu_{e})$
as function of the neutrino energy is shown for the $SO(10)$ \cite{SO10}, for 
$\sin^2(2 \theta_{e \tau}) =3.25 \times 10^{-10}$. Note that, the corresponding 
value of $E^{*}_{NA}$, not reported in the figure, results to be much
lower than $E_{A}$.}
\end{figure} 

\newpage

\begin{figure}[]
\epsfysize=18cm
\epsfxsize=16cm
\epsffile{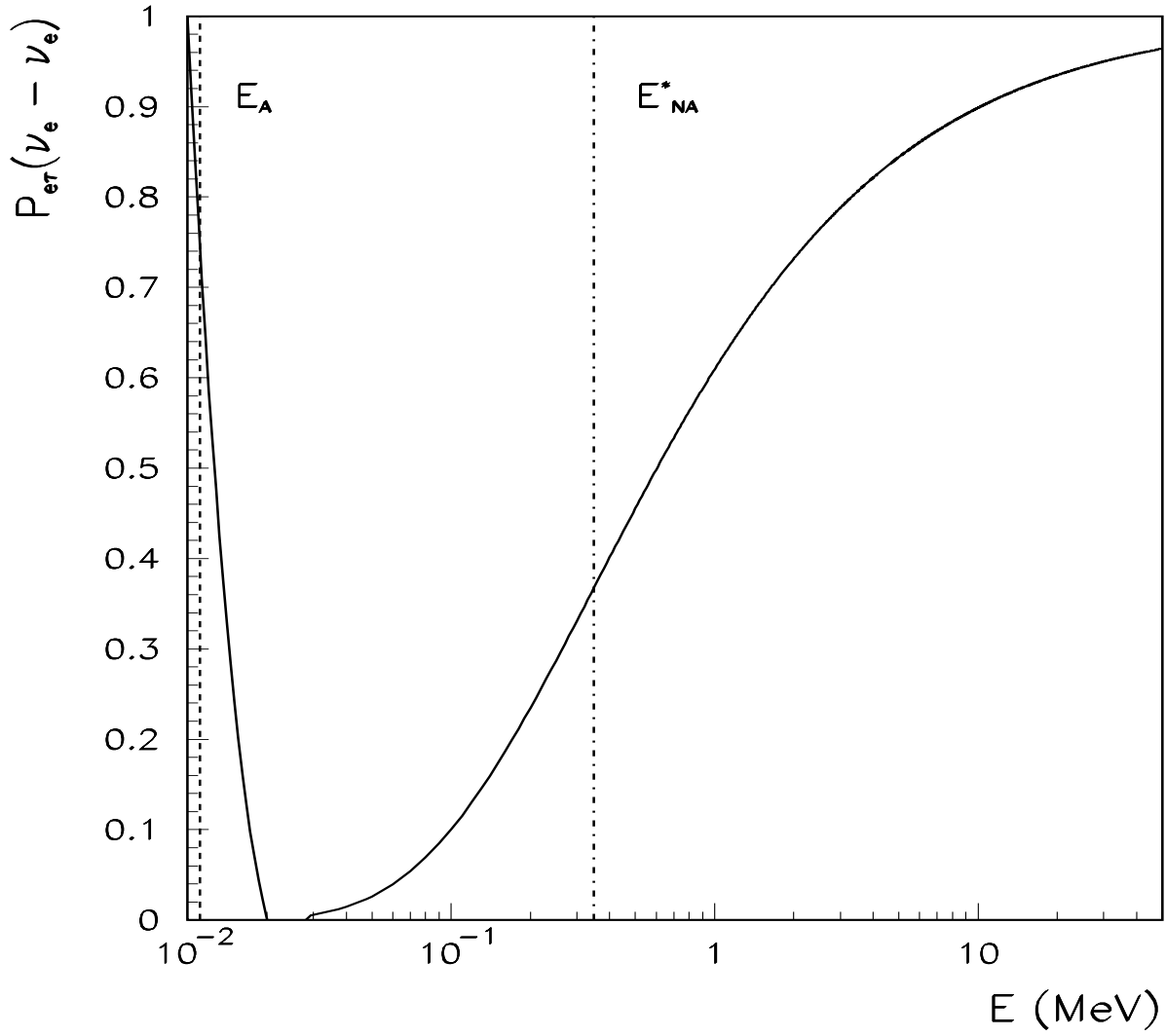}
\caption[]{The same quantity of Fig. 3 is shown, but with 
$\sin^2(2 \theta_{e \tau}) =3.25 \times 10^{-8}$.}
\end{figure} 

\newpage

\begin{figure}[]
\epsfysize=18cm
\epsfxsize=16cm
\epsffile{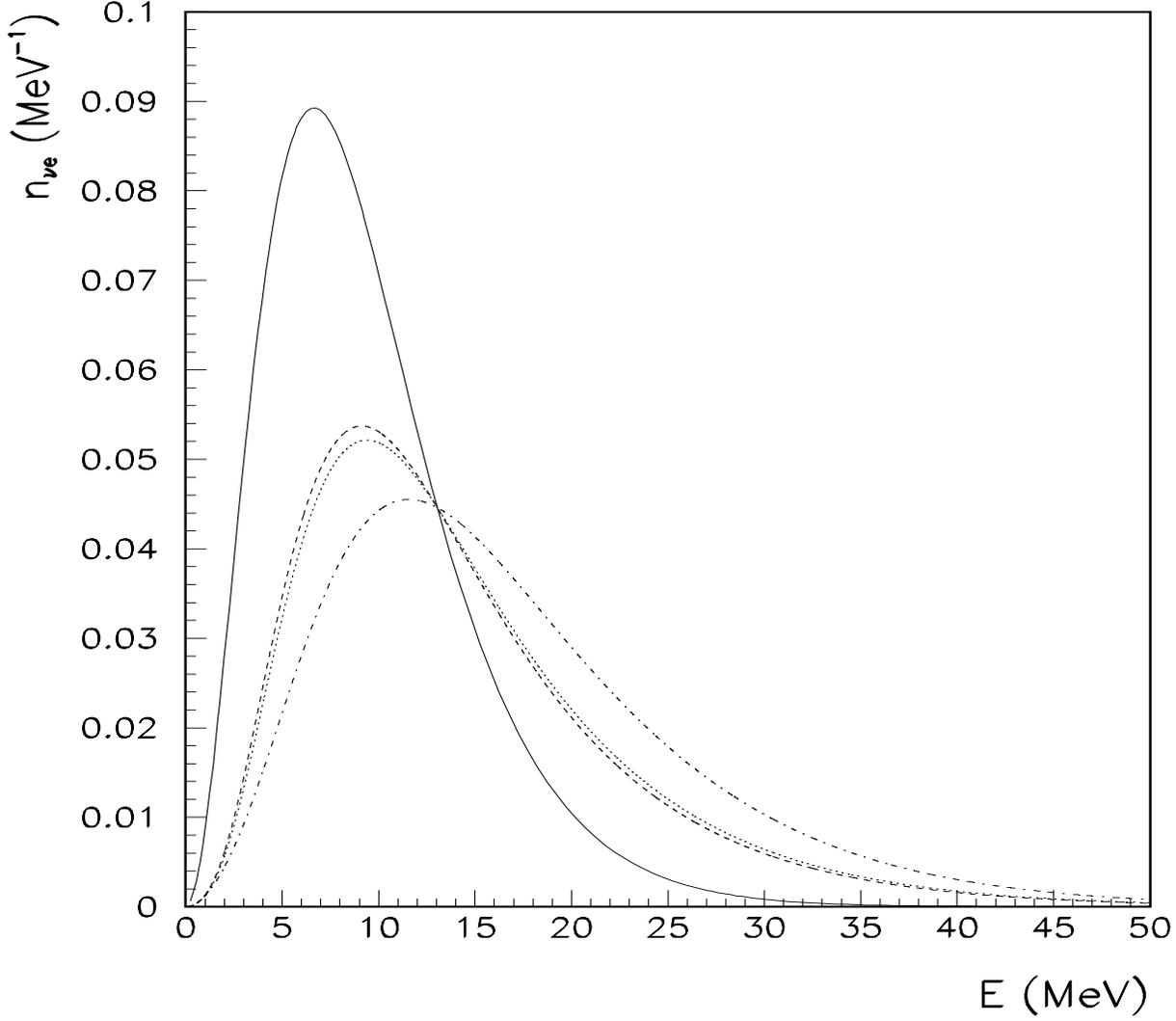}
\caption[]{The energy spectra for $\nu_{e}$ are here reported. The solid
line represents the pure initial thermal distribution at the neutrinosphere,
whilst the dashed and the dotted lines represent the distorted spectrum 
for $SO(10)$ \cite{SO10} with $\sin^2(2 \theta_{e \tau})=3.25 \times 10^{-10}$,
$3.25 \times 10^{-8}$, respectively. The dashed-dotted line corresponds to
the same quantity for $SU(5)_{SUSY}$, $SO(10)_{SUSY}$, in the DHR scheme
\cite{DHR}.}
\end{figure} 

\newpage

\begin{figure}[]
\epsfysize=18cm
\epsfxsize=16cm
\epsffile{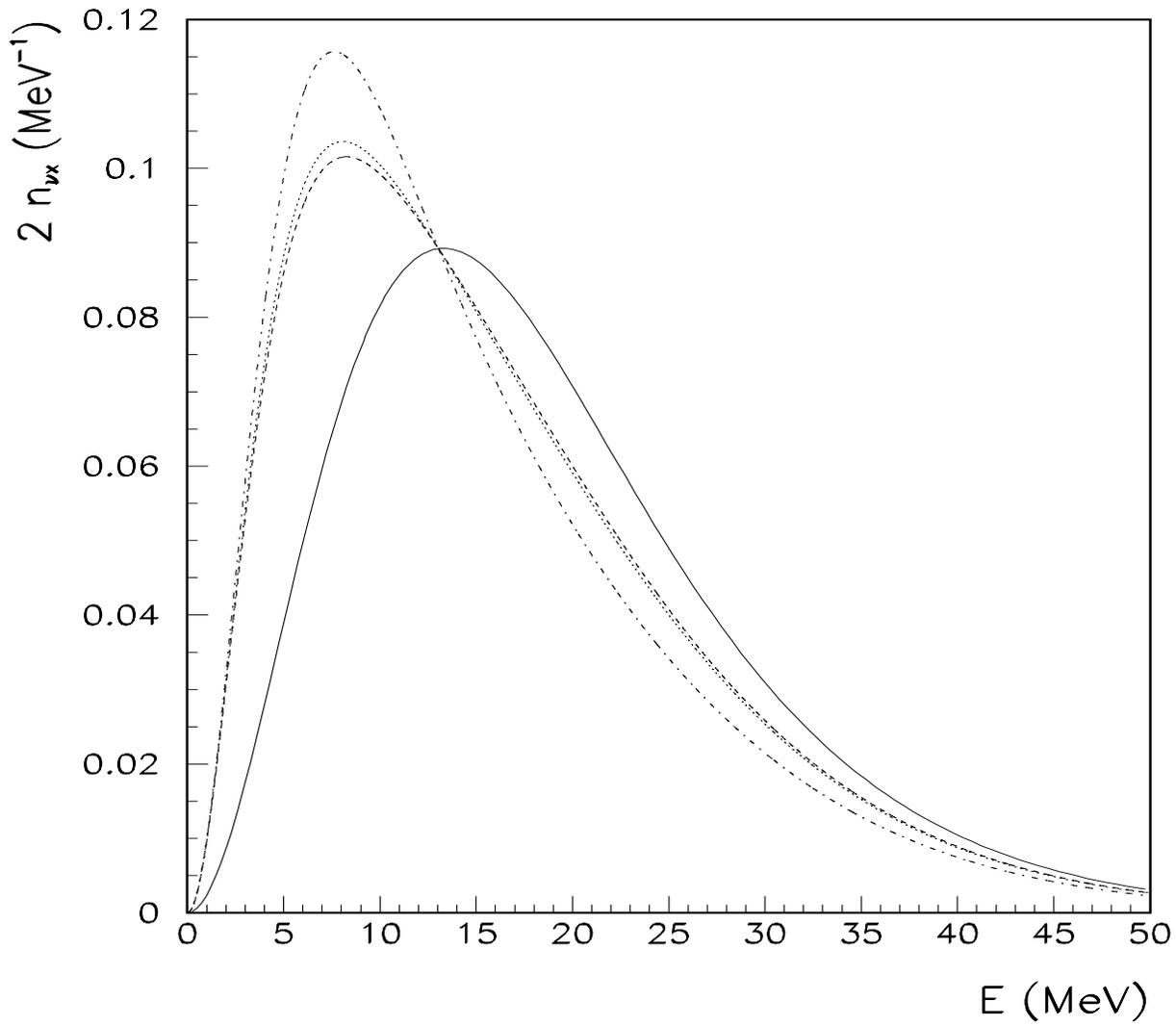}
\caption[]{The same quantities of Fig. 5 are here shown, but in this case 
corresponding to $\nu_{x}$.}
\end{figure} 
\end{document}